\documentclass[conference]{IEEEtran}
\IEEEoverridecommandlockouts
\usepackage{cite}
\usepackage{amsmath,amssymb,amsfonts,bm,dsfont}
\usepackage{graphicx}\graphicspath{{./figs/}}
\usepackage{subcaption}
\usepackage{tikz}
\usetikzlibrary{positioning}
\usepackage[colorlinks=false,hidelinks]{hyperref}
\usetikzlibrary{arrows,calc,chains,shapes.geometric,positioning,automata}

\DeclareMathOperator*{\argmax}{arg\,max}
\DeclareMathOperator*{\argmin}{arg\,min}

\def\BibTeX{{\rm B\kern-.05em{\sc i\kern-.025em b}\kern-.08em
    T\kern-.1667em\lower.7ex\hbox{E}\kern-.125emX}}
\begin{document}

\title{Tightly Integrated Motion Classification and State Estimation in Foot-Mounted Navigation Systems}

\author{\IEEEauthorblockN{Isaac Skog}
\IEEEauthorblockA{\textit{Dept. of Electrical Engineering} \\
\textit{Uppsala University}\\
Uppsala, Sweden \\
isaac.skog@angstrom.uu.se}
\and
\IEEEauthorblockN{Gustaf Hendeby}
\IEEEauthorblockA{\textit{Dept. of Electrical Engineering} \\
\textit{Link\"{o}ping University}\\
Link\"{o}ping, Sweden \\
gustaf.hendeby@liu.se}
\and
\IEEEauthorblockN{Manon Kok}
\IEEEauthorblockA{\textit{Delft Center for Systems and Control} \\
\textit{Delft University of Technology}\\
Delft, The Netherlands \\
m.kok-1@tudelft.nl}}

\maketitle

\begin{abstract}
A framework for tightly integrated motion mode classification and state estimation in motion-constrained inertial navigation systems is presented. The framework uses a jump Markov model  to describe the navigation system's motion mode and navigation state dynamics with a single model. A bank of Kalman filters is then used for joint inference of the navigation state and the motion mode. A method for learning unknown parameters in the jump Markov model, such as the motion mode transition probabilities, is also presented. The application of the proposed framework is illustrated via two examples. The first example is a foot-mounted navigation system that adapts its behavior to different gait speeds. The second example is a foot-mounted navigation system that detects when the user walks on flat ground and locks the vertical position estimate accordingly. Both examples show that the proposed framework provides significantly better position accuracy than a standard zero-velocity aided inertial navigation system. More importantly, the examples show that the proposed framework provides a theoretically well-grounded approach for developing new motion-constrained inertial navigation systems that can learn different motion patterns.
\end{abstract}


\bstctlcite{IEEEexample:BSTcontrol}

\section{Introduction}
\label{sec:intro}
Current state-of-the-art technology for zero-velocity aided inertial navigation systems is based upon a strategy of loose integration between the zero-velocity detector and the inertial navigation filter~\cite{Wahlstrom2021}. That is, the zero-velocity detector and the inertial navigation filter are treated as separate functions, with a one-directional flow of information from the former to the latter. From an information theoretical perspective this is suboptimal because: (a) the estimated navigation states carry information about the system's motion mode that is not used in the zero-velocity detector, and (b) the test statistic used in the zero-velocity detector is quantized before it is used in the navigation filter. Thus, information that could be used both to improve the detection of zero-velocity events and to control how the zero-velocity updates are performed, is lost. The same argumentation also applies to other motion constraints commonly applied to foot-mounted inertial navigation systems and where an external motion classifier is used to determine when to apply these constraints. A few examples of commonly used motion constraints are that the system keeps a constant height~\cite{Zhang2020}, constant heading~\cite{Abdulrahim2012}, or constant speed~\cite{Kronenwett2018}.

The fact that the estimated navigation state carries information that can be used to improve the zero-velocity detection process is utilized in~\cite{Wahlstrom2019}, where the velocity estimates from the inertial navigation filter are used as a prior for a Bayesian zero-velocity detector. And in~\cite{Jao2021}, the test statistics from the zero-velocity detector are used to control the magnitude of the measurement covariance in the inertial navigation filter. Thereby, quantization of the zero-velocity test statistics is avoided and the zero-velocity update process can adapt to different gait conditions. Still, both~\cite{Wahlstrom2019} and \cite{Jao2021} employ a loose integration strategy where the zero-velocity detector and the inertial navigation filter are treated as separate functions.

This paper instead proposes a framework for tightly integrated motion mode classification and navigation state estimation, of which tightly integrated zero-velocity aided inertial navigation is a special case. The core of the framework is a jump Markov model that includes both the navigation state and the motion mode. That is, a single model is used to describe both the kinematics of the system under various motion modes and the probability of transitioning between the motion modes. A bank of Kalman filters is then used to jointly estimate the navigation state and the motion mode. A method to automatically learn unknown parameters in the jump Markov model  is also presented. The proposed framework is evaluated on two data sets consisting of various gait conditions and motion modes.\\[-0.5em] 

\noindent\textbf{Reproducible research:} The data and code used to produce the presented results can be downloaded at:\\\rule{0pt}{0pt}\quad\texttt{\small\url{https://gitlab.liu.se/open-shoe/filterbanks}}.


\section{Signal Model}\label{S:prob formulation}
Let the system state $x_k$ and input vector $u_k$ at time instant $k$ be defined as
\begin{equation}
  x_k=
  \begin{bmatrix}
    r_k\\ v_k\\ q_k\\ \xi_k\end{bmatrix}\quad \text{and}\quad
  u_k= \begin{bmatrix}
    s_k\\ \omega_k
  \end{bmatrix},
\end{equation}
respectively. Here $r_k$, $v_k$, and $q_k$ denote the position, velocity, and attitude quaternion, respectively. Further, $\xi_k$ denotes potential auxiliary states, such as sensor biases, needed to describe the behavior of the system. Moreover, $s_k$ and $\omega_k$ denote the specific force and angular velocity, respectively.

Next, define the discrete state
\begin{equation}
  \delta_k\in\{1,\ldots, L\}
\end{equation}
that indicates the current motion mode of the navigation system, e.g., the system is stationary, keeping a constant height, or moving at constant velocity. A jump Markov nonlinear model that can be used to the describe the dynamics of the inertial navigation system is then given by~\cite{Mazor1998}
\begin{subequations}\label{eq:jmns}
  \begin{align}
    &x_{k+1}=f(x_k,\delta_k,u_k,\eta_k)\\
    &y_k=h(x_k,\delta_k,u_k)+e_k\\
    &\eta_k\overset{\text{\tiny i.i.d.}}{\sim}\mathcal{N}\bigl(n_k;0,Q_k\bigr)\\
    &e_k\overset{\text{\tiny i.i.d.}}{\sim}\mathcal{N}\bigl(e_k;0,R(\delta_k)\bigr)\\
    &p(\delta_{k+1}|\delta_k)=\Pi_{\delta_{k+1},\delta_{k}}.
  \end{align}
\end{subequations}
%
Here $y_k\triangleq 0$  is used as a pseudo-observation to impose a set of motion mode dependent ``stochastic constraints'', such as zero-velocity constraints, on the navigation state $x_k$. The exact form of these constraints, defined by $h(\cdot)$, depends on the assumed motion modes and will be discussed later. The ``hardness'' of the imposed constraints is controlled via the motion mode dependent covariance $R(\delta_k)$. Further, $\mathcal{N}\bigl(\,\cdot\,;\mu,\Sigma\bigr)$ denotes a multivariate normal distribution parameterized by the mean vector $\mu$ and covariance matrix $\Sigma$. The function $f(\cdot)$ describes the system dynamics and is given by the inertial navigation equations and the dynamics of the auxiliary states~$\xi_k$. Moreover, $Q_k$ denotes the covariance of the process noise, and  $\Pi_{i,j}$ denotes the $i$:th row and $j$:th column entry of the mode transition probability matrix. Finally, $p(a|b)$ denotes the probability density (mass) function of $a$ given $b$. Note that it is straightforward to extend the model (\ref{eq:jmns}) to also include real observations from various sensors, but it is out of the scope of this paper, as the focus is on motion-constrained inertial navigation.


\section{State Estimation Using a Filter Bank}\label{sec:solutions}
A variety of inference techniques for jump Markov models exist. Here a commonly used filter banks solution to the inference problem will be recapitulated; for details the reader is referred to~\cite{Blom1998} and \cite{Rong2005}. Thereafter, necessary approximations needed to adapt the filter bank solution to the considered navigation problem will be presented.

\subsection{Linear model with normal distributed noise}
The goal of the inference process is to estimate the a posteriori distribution $p(x_k|y_{1:k})$ of the state $x_k$ given all the observations up until time~$k$, denoted as~$y_{1:k}$. If the motion mode sequence $\delta_{1:k}$ is known, the state-space model in the Markov jump system is linear, and the process and measurement noise are normally distributed, then the a posteriori distribution can be estimated with the Kalman filter. That is,
\begin{equation}\label{eq:a posteriori}
    p(x_k|\delta_{1:k},y_{1:k})=\mathcal{N}\bigl(x_k;\hat{x}^{\delta_{1:k}}_{k|k},P^{\delta_{1:k}}_{k|k}\bigr),
\end{equation}
where $\hat{x}_{k|k}^{\delta_{1:k}}$ and $P_{k|k}^{\delta_{1:k}}$ denote the Kalman filter state estimate and state covariance given the mode sequence $\delta_{1:k}$, respectively. In reality, the motion mode sequence is unknown and must be estimated from the measurements. Let $p(\delta_{1:k}|y_{1:k})$ denote the a posteriori distribution of the motion mode sequences $\delta_{1:k}$ given the measurements $y_{1:k}$. The sought-after a posteriori distribution of the state $x_k$ can then be found as
\begin{equation}
\begin{split}
    p(x_k|y_{1:k})&=\sum_{i=1}^{L^k}p(x_k|\delta^{i}_{1:k},y_{1:k})p(\delta^{i}_{1:k}|y_{1:k})\\
     & =\sum_{i=1}^{L^k}w^i_k\mathcal{N}\bigl(x_k;\hat{x}^{\delta^i_{1:k}}_{k|k},P^{\delta^i_{1:k}}_{k|k}\bigr),
\end{split}
\end{equation}
where $w^i_k\triangleq p(\delta^{i}_{1:k}|y_{1:k})$ is probability of the $i$:th motion mode sequence. Hence, the posteriori distribution of the state $x_k$ is given by a mixture of $L^k$ weighted normal distributions where the expected value and covariance of each mixture component are calculated via a Kalman filter. That is, the posteriori distribution can be calculated via a bank of Kalman filters where the size of the filter bank grows according to a tree structure. The weights $w^i_k$ in the mixture, i.e., the probability for each branch in the tree, can be recursively calculated as
\begin{subequations}
\begin{equation}
  w^i_k\propto p(y_k|y_{1:k-1},\delta^i_{1:k})p(\delta^i_k|\delta^i_{k-1})w^i_{k-1},
\end{equation}
where
\begin{equation}
  \sum_{i=1}^{L^k}w^i_k=1,
\end{equation}
and
\begin{equation}\label{eq:normal distribution likelihood}
p(y_k|y_{1:k-1},\delta^i_{1:k})=\mathcal{N}\bigl(y_k;\hat{y}_{k|k-1}^{\delta^i_{1:k}},S_k^{\delta^i_{1:k}}\bigr).
\end{equation}
\end{subequations}
Here $\hat{y}_{k|k-1}^{\delta_{1:k}}$ and $S_k^{\delta_{1:k}}$ denote the Kalman filter measurement prediction and innovation covariance, respectively.

From the posteriori distribution, the minimum variance estimate of the state $x_k$ can be calculated as
\begin{subequations}
\begin{equation}\label{eq:mv}
  \hat{x}^{\text{\scriptsize mv}}_k=\sum_{i=1}^{L^k}w^{i}_k \hat{x}_{k|k}^{\delta^i_{1:k}}
\end{equation}
and the associated conditional covariance matrix as
\begin{equation}
   P^{\text{\scriptsize mv}}_k=\sum_{i=1}^{L^k}w^{i}_k \bigl(P_{k|k}^{\delta^i_{1:k}}+(\hat{x}_{k|k}^{\delta^i_{1:k}}-\hat{x}^{\text{\scriptsize mv}}_k)(\hat{x}_{k|k}^{\delta^i_{1:k}})-\hat{x}^{\text{\scriptsize mv}}_k)^\top\bigr).
\end{equation}
\end{subequations}

\subsection{Approximative solution}
The outlined filter bank solution to the inference problem can in general not be used without modifications due to the growing number of Kalman filters needed. Therefore, many strategies have been developed for pruning and merging branches in the growing filter bank tree so that a fixed complexity is achieved~\cite{Li1996}.

Beyond the challenges with the exponentially increasing complexity, the outlined general solution cannot still be applied to the Markov jump system model in~\eqref{eq:jmns}. The system is nonlinear and the attitude states belong to a manifold of Euclidean space. This implies that $p(x_k|\delta_{1:k},y_{1:k})$ in \eqref{eq:a posteriori} is generally not a normal distribution, neither can the minimum variance estimate of the attitude $q_k$ be calculated as in \eqref{eq:mv}. A common way to handle that the system is nonlinear and the attitude not belonging to Euclidian space in the filtering process is to use an error state Kalman filter~\cite{Sola2017}. The a posteriori distribution $p(x_k|\delta_{1:k},y_{1:k})$ is then, at every time instant, approximated as normal distributed with the mean and covariance given by the error state Kalman filter~\cite{Kok2017}.

One way to calculate a point estimate of the attitude in terms of the Euler angles $\phi$ is via
\begin{equation}\label{eq:mv attitude}
  \hat{\phi}^{\text{\scriptsize mv}}_k=\argmin_{\phi\in\Omega_{\phi}}\sum_{i}^{L^k}w^i_k\|C(\hat{q}^{\delta^i_{1:k}}_{k|k})-C(\phi)\|^2_F,
\end{equation}
where $C(q)$ denotes the rotation matrix that transforms a vector from the body frame to the navigation frame~\cite{Moakher2002}; to simplify the notation, a slight misusage of notation is admitted, and the rotation matrix is here interchangeably parameterized by the attitude quaternion $q$ and the corresponding Euler angles $\phi$.  Further, the set $\Omega_{\phi}=[0,2\pi) \times [-\pi/2,\pi/2) \times [0,2\pi)$. The covariance of $\hat{\phi}^{\text{\scriptsize mv}}_k$ can be approximately calculated as
\begin{equation}
  \text{Cov}(\hat{\phi}^{\text{\scriptsize mv}}_k)\approx\sum_{i}^{L^k}w^i_k\bigl(\Sigma^{\delta^i_{1:k}}_{k|k}+\Delta\phi^{\delta^i_{1:k}}_k(\Delta\phi^{\delta^i_{1:k}}_k)^\top\bigr).
\end{equation}
Here $\Sigma^{\delta^i_{1:k}}_{k|k}$ denotes the covariance of the attitude (in terms of Euler angles) calculated by the error state Kalman filter for the branch corresponding to the motion sequence $\delta^i_{1:k}$. Further, the difference between the attitude estimate calculated by the error state Kalman filter and the minimum variance attitude estimate in \eqref{eq:mv attitude} is given by
\begin{equation}
  \Delta\phi^{\delta^i_{1:k}}_{k|k}=\{\phi\in\Omega_{\phi}~\text{s.t.}~~C(\hat{\phi}^{\text{\scriptsize mv}}_k)=C(\phi)C(\hat{q}^{\delta^i_{1:k}}_{k|k})\}.
\end{equation}
Note that the optimization problem in (\ref{eq:mv attitude}) can be efficiently solved using a polar decomposition \cite{Moakher2002}.


\section{Learning of Unknown Model Parameters}\label{sec:learn}
The signal model in \eqref{eq:jmns} may have unknown parameters, such as the transition probability matrix $\Pi$, whose values can be hard to specify accurately. Instead of resorting to cumbersome hand-tuning of these parameters, they may be learned from data. A variety of methods to learn model parameters in jump Markov models has been suggested~\cite{Jilkov2004,Ozkan2015}. Here a maximum likelihood method, similar to that presented in \cite{Kok2017}, for learning the parameters from the data will be outlined.

Let $\theta$ denote the unknown parameters in the model. The maximum likelihood estimate of these parameters is then given by
\begin{subequations}
\begin{equation}\label{eq:learn theta}
  \hat{\theta}=\argmax_\theta p(y_{1:k};\theta),
\end{equation}
where
\begin{equation}
  p(y_{1:k};\theta)=\prod_{n=2}^{k}p(y_n|y_{1:n-1})p(y_1),
\end{equation}
and
\begin{equation}\label{eq:pyyn}
    p(y_n|y_{1:n-1})=\sum_{i=1}^{L^n}\mathcal{N}\bigl(y_n;\hat{y}_{n|n-1}^{\delta^i_{1:n}},S_n^{\delta^i_{1:n}}\bigr)p(\delta^i_n|\delta^i_{n-1})w^i_{n-1}.
\end{equation}
\end{subequations}
Here $p(y_1)$ denotes the a priori probability of the observation~$y_1$. Note that the number of modes in the full mixture distribution, which subsequently must be summed to marginalize away the mode dependence, increases exponentially in $n$.  However, in practice, the actual number of modes to consider is reduced using pruning or merging in the filter bank.  Hence, the number of modes to consider when evaluating \eqref{eq:pyyn} can be kept tractable.


\section{Application Examples}
Next, the proposed inference framework is used to realize two foot-mounted inertial navigation systems that incorporate different motion modes. The first system adaptively selects the covariance matrix used in the zero-velocity update, and the second system automatically detects if the system returns to the same height after a step. For both systems, the motion mode state transition matrix is learned from training data using the method outlined in Sec.~\ref{sec:learn}. The performances of both systems are compared to the OpenShoe system presented in~\cite{Nilsson2014}.

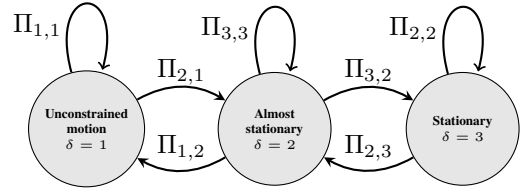
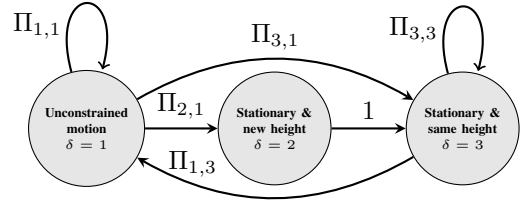
\begin{figure}[t!]
\begin{subfigure}{\columnwidth}
\centering
   \hrule
\begin{tikzpicture}[node distance = 2.5cm, on grid, every state/.style = {draw = black, fill = gray!20,font=\tiny,minimum size=1.5cm},auto]
    \node (M) [state,align=center] at (0,0) {\textbf{Unconstrained}\\ \textbf{motion}  \\$\delta=1$}; 

    \node (AS) [state, right = of M,align=center] {\textbf{Almost} \\ \textbf{stationary}\\$\delta=2$};

    \node (S) [state,right = of AS, align=center] {\textbf{Stationary}\\$\delta=3$}; 

\path [-stealth, thick]
    (M) edge [loop above]  node [left,xshift=-0.2cm,yshift=-0.3cm] {$\Pi_{1,1}$}()
    (S) edge [loop above]  node [left,xshift=-0.2cm,yshift=-0.3cm] {$\Pi_{2,2}$}()
    (AS) edge [loop above]  node [left,xshift=-0.2cm,yshift=-0.3cm] {$\Pi_{3,3}$}()
    (M) edge [bend left] node [above,yshift=-0.1cm] {$\Pi_{2,1}$} (AS)
    (AS) edge [bend left]  node [above,yshift=-0.1cm] {$\Pi_{3,2}$}(S)
    (S) edge [bend left] node [above] {$\Pi_{2,3}$} (AS)
    (AS) edge [bend left]  node [above] {$\Pi_{1,2}$}(M);
\end{tikzpicture}
\caption{Varying gait speed example.}\label{fig:state transition varying gait speed}
\hrule
\end{subfigure}
\begin{subfigure}{\columnwidth}
\centering
\begin{tikzpicture}[node distance = 2.5cm, on grid, every state/.style = {draw = black, fill = gray!20,font=\tiny,minimum size=1.5cm},auto]
    \node (M) [state,align=center] at (0,0) {\textbf{Unconstrained}\\ \textbf{motion} \\$\delta=1$};

    \node (S) [state, right = of M,align=center] {\textbf{Stationary \&}\\ \textbf{new height}\\$\delta=2$};

    \node (SSH) [state,right = of S, align=center] {\textbf{Stationary \&}\\ \textbf{same height}\\$\delta=3$};

\path [-stealth, thick]
    (M) edge [loop above]  node [left,xshift=-0.2cm,yshift=-0.3cm] {$\Pi_{1,1}$}()
    (M) edge [bend left] node [above] {$\Pi_{3,1}$} (SSH)
    (SSH) edge [loop above]  node [left,xshift=-0.2cm,yshift=-0.3cm] {$\Pi_{3,3}$}(S)
    (M) edge node [above,xshift=0cm] {$\Pi_{2,1}$} (S)
    (S) edge node [above,xshift=0cm] {$1$} (SSH)
    (SSH) edge [bend left] node [above,xshift=-1.1cm,yshift=0.15cm] {$\Pi_{1,3}$} (M);
       \vspace{-1mm}
\end{tikzpicture}
\caption{Return to same height example.}\label{fig:state transition stair walking}
   \hrule
\end{subfigure}
\caption{Mode transition diagrams for the application examples.}\label{fig:state transition diagram}
\end{figure}

%

\subsection{Example: Varying gait speed}\label{sec:mixedgait}
The challenge of designing and tuning a foot-mounted zero-velocity aided inertial navigation system so that it works well for multiple gait speeds is well-known~\cite{Wahlstrom2021}. To use the proposed framework to design a system that adaptively selects the detection threshold, as well as the covariance matrix used in the zero-velocity updates, we define the following three motion modes: the unconstrained motion mode ($\delta_k=1$), the almost stationary motion mode ($\delta_k=2$), and the stationary motion mode ($\delta_k=3$). A motion mode transition diagram illustrating how the system is assumed to transition between these motion modes is shown in Fig.~\ref{fig:state transition varying gait speed}. The associated motion mode transition probabilities $\Pi_{i,j}$ are also shown.

Next, introduce the following system dynamics
\begin{subequations}\label{eq:varying gait}
\begin{equation}\label{eq:f varying gait}
  f(x_k,\delta_k,u_k,\eta_k)=\begin{bmatrix}
                           r_k+\Delta tv_k\\
                            v_k+\Delta t \bigl(C(q_k)(s_k+\eta^{s}_k)+g\bigr)\\
                           \Omega\bigl(\Delta t\,(\omega_k+\eta^{\omega}_k)\bigr)q_k\\
                         \end{bmatrix}.
\end{equation}
Here $\Delta t$ and $g$ denote the sampling period and gravity vector, respectively. Further, $\Omega(\cdot)$ denotes the quaternion update matrix (see~\cite{Farrell1998} for details). Moreover, $\eta^{s}_k$ and $\eta^{\omega}_k$ denote the process noise associated with the accelerometers and gyroscopes, respectively.

Finally, define the stochastic motion constraints imposed during the different motion modes as follows
\begin{equation}\label{eq:h varying gait}
  h(x_k, u_k,\delta_k)=\begin{cases}
                         \hfil~~~0 \hfil, & \delta_k=1 \\
                         h_0(x_k,u_k), & \delta_k=\{2,3\}
                       \end{cases},
\end{equation}
where
\begin{equation}\label{eq:h0}
  h_0(x_k,u_k)=\begin{bmatrix}
                                     v_k \\
                                     \omega_k\\
                                     C(q_k)s_k+g
                                  \end{bmatrix}.
\end{equation}
The ``hardness'' of the constraints is controlled by the covariance matrix
\begin{equation}\label{eq:R varying gait}
   R(\delta_k)=\begin{cases}
                 \sigma^2_{nc} I, & \delta_k=1 \\
                 R_0(\delta_k), & \delta_k=\{2,3\}
               \end{cases},
\end{equation}
where
\begin{equation}\label{eq:R0}
  R_0(\delta_k)=\sigma^2_v({\delta_k})I \oplus \sigma^2_\omega(\delta_k)I \oplus \sigma^2_s(\delta_k)I.
\end{equation}
\end{subequations}
Here $\oplus$ denotes the direct sum matrix operator and $I$ denotes an identity matrix of appropriate size. Further, $\sigma^2_v({\delta_k})$, $\sigma^2_\omega(\delta_k)$, and $\sigma^2_s(\delta_k)$ denote the mode-dependent variance associated with the stochastic constraints on the velocity, angular velocity, and acceleration, respectively. Moreover, $\sigma^2_{nc}$ is a design parameter that controls the measurement likelihood associated with no constraints. 

A filter bank with the jump Markov model defined by \eqref{eq:varying gait} was designed and two data sets were recorded with a sensor array consisting of 32 InvenSense MPU9150 IMUs while a person walked and ran back and forth along a straight line. From the recorded data, two data sets with 50 measurement sequences each were created by repeatedly drawing four random IMUs and averaging their measurements. These measurement sequences were then processed using the OpenShoe system algorithm with the zero-velocity detector threshold tuned to generate a detection at least once per gait cycle. The measurements were also processed using the designed filter bank. Pruning was used to keep the tree size to a maximum of nine leaves, and a pruning strategy where the most probable leaves were retained was used.

The measurement sequences in the first data set were used to learn the mode transition matrix $\Pi$. The initial value and end result of the learning process were
\begin{equation}
\setlength\arraycolsep{3pt}
  \widehat{\Pi}^{\text{\tiny init}}=\begin{bmatrix}
        1/2 & 1/3 & 0 \\
        1/2 & 1/3 & 1/2 \\
        0 & 1/3 & 1/2
      \end{bmatrix}~\text{and}~~\widehat{\Pi}^{\text{\tiny end}}=\begin{bmatrix}
     0.993  &  0.073  &  0\\
     0.007  &  0.893  &  0.005\\
      0 &  0.034  & 0.995\\
      \end{bmatrix},\nonumber
\end{equation}
respectively, and the optimization converged in less than 10 iterations. The matrix $\widehat{\Pi}^{\text{\tiny end}}$  corresponds to a mode transition system with low-pass characteristics. That is, the probability of staying in a mode is much higher than transitioning to another mode. On average, the percentage of time spent in the motion modes one, two, and three is 57\%, 5\%, and 38\%, respectively.

\begin{figure}[t]
  \centering
  \begin{subfigure}{.94\columnwidth}
\centering
  \includegraphics[trim={4.2cm 9.3cm 4.7cm 9.7cm},clip,width=\columnwidth]{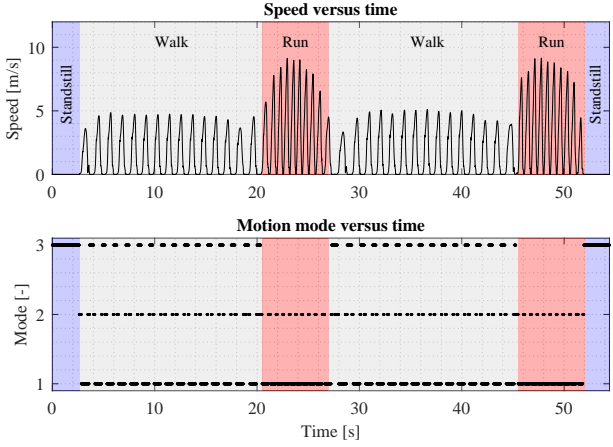}
  \caption{Estimated speed and most likely motion mode versus time.}\label{fig:speedvstime}
  \end{subfigure}
  \par\bigskip
   \begin{subfigure}{.94\columnwidth}
  \centering
  \includegraphics[trim={4.2cm 9.3cm 4.7cm 9.7cm},clip,width=\columnwidth]{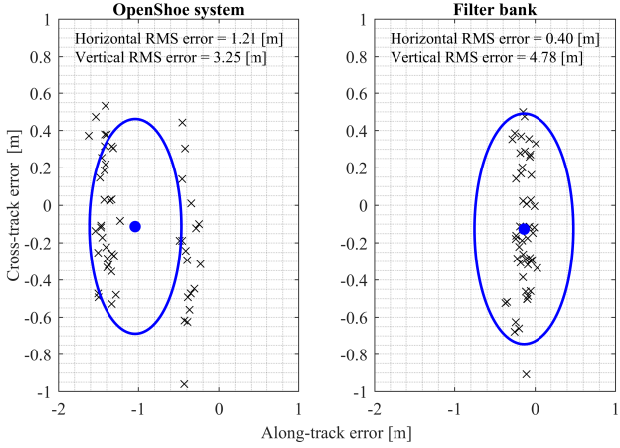}
  \caption{Horizontal position error at end of the trajectory. Also shown are the mean error (blue dot) and 95\% confidence interval (blue circle) calculated from the filter covariance, assuming the error to be normally distributed.}\label{fig:poserror}
 \end{subfigure}
 \caption{Results from varying gait speed application example.}\label{fig:varyingspeed}
\end{figure}

The learned mode transition matrix was then used in the filter bank when processing the measurements in the second data set. The result is shown in Fig.~\ref{fig:varyingspeed}. In Fig.~\ref{fig:speedvstime} the speed and most likely mode sequence $\hat{\delta}_{1:k}$ estimated by the filter bank from one of the measurement sequences are shown. And in Fig.~\ref{fig:poserror} the horizontal plane errors at the end of the trajectory are shown.

From the figures, the following things can be observed. The filter bank adaptively selects different motion modes depending on the gait speed. That is, when walking, the filter bank frequently selects the stationary motion mode as the most likely mode, whereas when running, the filter bank never selects the stationary mode. Comparing the horizontal positioning error of the OpenShoe system and the filter bank the systems have about the same cross-track error, whereas the filter bank has a significantly smaller along-track error. The OpenShoe system has an along-track bias error of about minus one meter. This is likely due to the high detection threshold used in the zero-velocity detector, causing zero-velocity updates to be applied even when the system is moving. This, in turn, causes part of each step to be cut away, especially when the user walks and the foot's transition from stationary to moving is less distinct. Thanks to the ability of the filter bank to adaptively select the detection threshold and the covariance used in the zero-velocity update, this effect is reduced. However, looking at the vertical root mean square (RMS) error, both systems have an error of several meters. How to reduce this error will be illustrated next.

\begin{figure}[t]
  \centering
  \begin{subfigure}{.94\columnwidth}
\centering
   \includegraphics[trim={4.2cm 9.3cm 4.7cm 9.7cm},clip,width=\columnwidth]{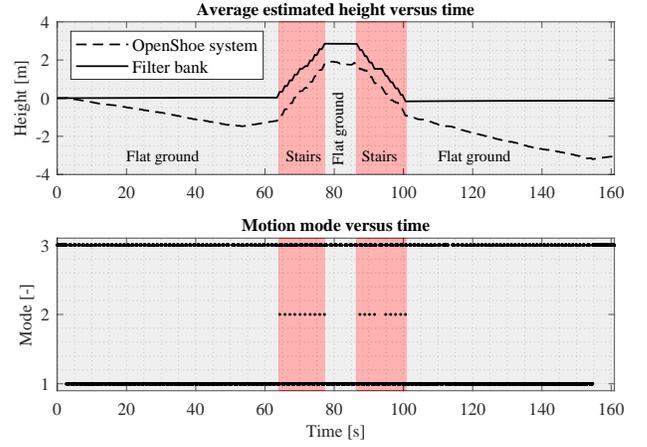}
  \caption{Estimated height and most likely motion mode versus time.}\label{fig:heightvstime}
  \end{subfigure}
  \par\bigskip
   \begin{subfigure}{.94\columnwidth}
  \centering
  \includegraphics[trim={4.2cm 9.3cm 4.7cm 9.7cm},clip,width=\columnwidth]{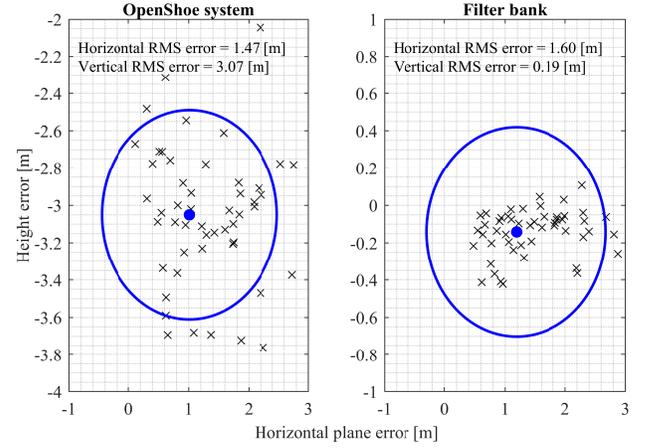}
  \caption{Height versus horizontal position error at end of the trajectory. Also shown are the mean error (blue dot) and 95\% confidence interval (blue circle) calculated from the filter covariance, assuming the error to be normally distributed.}\label{fig:poserror2}
 \end{subfigure}
 \caption{Results from return to same height application example.}\label{fig:stair walking}
\end{figure}

\subsection{Example: Return to same height}\label{sec:stairwalking}
The vertical position error often grows faster than the horizontal position error in foot-mounted zero-velocity aided inertial navigation systems. This is because the beginning and end of each step, when the foot is mainly moving in the vertical direction, are cut away by the zero-velocity updates. However, the vertical error growth can be significantly reduced by noting that most of the time a person is walking on flat ground and the foot should return to the same height~\cite{Abdulrahim2012}. To design a system using the proposed framework that incorporates this information, we introduce the following three motion modes: the unconstrained motion mode ($\delta_k=1$), the stationary at new height motion mode ($\delta_k=2$), and the stationary at the same height motion mode ($\delta_k=3$). A motion mode transition diagram illustrating how the system is assumed to transition between these motion modes is shown in Fig.~\ref{fig:state transition stair walking}. Here it has been assumed that if the system becomes stationary at a new height, the system will be stationary for at least two samples. Hence the probability of transitioning from mode $\delta_k=2$ to mode $\delta_k=3$ is one.

Next, we introduce the following extended system dynamics
\begin{subequations}
\begin{equation}\label{eq:f same height}
  f_{\text{\tiny ext}}(x_k,\delta_k,u_k,\eta_k)=\!\begin{bmatrix}
                           f(x_k,\delta_k,u_k,\eta_k)\\
                           \mathds{1}(\delta_k=1)\xi_k+\mathds{1}(\delta_k \neq 1)[p_k]_3\\
                         \end{bmatrix},
\end{equation}
where $\mathds{1}(\cdot)$ denotes the indicator function and $[a]_j$ denotes $j$:th element of the vector $a$. Furthermore, the auxiliary state $\xi_k$ stores the height from the time when the system was last stationary.

The stochastic motion constraints are as follows
\begin{equation}\label{eq:h same height}
   h_{\text{\tiny ext}}(x_k, u_k,\delta_k)=\begin{cases}
                                \hfil~~~0 \hfil, & \delta_k=1\\
                                 ~~h_0(x_k, u_k), & \delta_k=2 \\
                                 \begin{bmatrix}
                                  h_0(x_k, u_k)\\
                                     [p_k]_3-\xi_k
                                  \end{bmatrix}, &\delta_k=3 \\
                               \end{cases}.
\end{equation}
The ``hardness''  of these constraints is controlled by the covariance matrices
\begin{equation}\label{eq:R same height}
    R_{\text{\tiny ext}}(\delta_k)=\begin{cases}
    \sigma^2_{nc} I, & \delta_k=1\\
    R_0(\delta_k),&\delta_k=2\\
    R_0(\delta_k) \oplus \sigma^2_h(\delta_k) I,&\delta_k=3\\
                               \end{cases},
\end{equation}
\end{subequations}
where $\sigma^2_h(\delta_k)$ denotes the variance associated with the stochastic constraints on height.

The jump Markov model defined by \eqref{eq:f same height}, \eqref{eq:h same height}, and \eqref{eq:R same height} was used to design a filter bank, and two data recordings were collected while a person walked on flat ground and then climbed and descended a stair. Following the same procedure as in the previous example, the data recordings were used to create two data sets with 50 measurement sequences each. The measurement sequences in the first data set were used to learn the mode transition matrix $\Pi$. The initial value and end result of the learning process were
\begin{equation}
\setlength\arraycolsep{3pt}
  \widehat{\Pi}^{\text{\tiny init}}=\begin{bmatrix}
        1/3 & 0 & 1/2 \\
        1/3 & 0 & 0 \\
        1/3 & 1 & 1/2
      \end{bmatrix}~\text{and}~~\widehat{\Pi}^{\text{\tiny end}}=\begin{bmatrix}
        0.976 & 0 & 0.031 \\
        0.003 & 0 & 0 \\
        0.021 & 1 & 0.969
      \end{bmatrix},\nonumber
\end{equation}
respectively, and the optimization converged in less than 10 iterations. Once again the matrix corresponds to a mode transition system with low-pass characteristics, where the average percentages of time spent in motion modes one, two, and three is 56.2\%, 0.2\%, and 43.6\%, respectively.

The learned mode transition matrix was then used in the filter bank when processing the measurements in the second data set. Once again, pruning was used to keep the tree size to a maximum of nine leaves and a pruning strategy where the most leaves were retained was used. The result is shown in Fig.~\ref{fig:stair walking}. In Fig.~\ref{fig:heightvstime} the height and the most likely mode sequence $\hat{\delta}_{1:k}$ estimated by the filter bank is shown. And in Fig.~\ref{fig:poserror2} the height versus horizontal plane errors at the end of the trajectory are shown.

From the figures, the following things can be observed. The filter bank can detect whether the user walks on flat ground or not, i.e., whether the foot returns to the same height as in the previous step or not. Therefore it effectively reduces the height error from several meters to a few decimeters. A minor increase in the horizontal position error is observed with the filter bank.


\section{Discussion and Conclusions}
A framework for tightly integrated motion classification and state estimation in motion-constrained inertial navigation systems has been presented. The framework provides a structured and theoretically sound way to design motion-constrained inertial navigation systems that can learn different motion patterns. The application of the framework has been illustrated via two examples of foot-mounted zero-velocity aided inertial navigation systems.  The examples show that a significant performance gain can be achieved compared to a standard foot-mounted inertial navigation system. However, the price paid is an increased computational complexity (proportional to the number of filters in the filter bank) and an increased number of system parameters to tune. The challenges with tuning the system parameters can to some extent be alleviated via the proposed parameter learning method. Still, tuning the system parameters can be challenging, especially if many motion modes are included in the system model.

\section{Outlook and Future Research}
A basic version of the filter bank framework has been presented and tested with models that have a few motion modes. Many possible improvements and open research questions exist. First and foremost, the framework should be compared against other adaptive zero-velocity detector frameworks. More complex models that include a variety of motion modes should also be explored. Related to that, hieratical model structures constructed of several linked small hidden Markov models for the motion states are of special interest to obtain a computationally attractive algorithm. Further, to obtain a smoother transition between motion modes and a more robust algorithm framework, the feasibility of substituting the normal distribution in (\ref{eq:normal distribution likelihood}) with a more heavy-tailed distribution, such as the Student-t distribution, should be explored. Moreover, since the probability of transitioning between motion modes varies both with time and the motion dynamics, the use of a constant mode transition probability matrix $\Pi$ is clearly suboptimal. Therefore, the feasibility of learning a time-varying and motion-dynamic dependent transition matrix should be explored.



\section*{Acknowledgment}
This work has been partially funded by the Swedish Research Council project 2020-04253 \emph{Tensor-field based localization} and the Dutch Research Council (NWO) research program Veni project 18213 \emph{Sensor Fusion For Indoor Localisation Using The Magnetic Field}.

\IEEEtriggeratref{99}
\bibliographystyle{IEEEtran}
\bibliography{IEEEabrv,magnetic_field_ref}

\end{document}